# Ultrafast carrier dynamics in thin-films of the topological insulator $Bi_2Se_3$

Yuri D. Glinka, Sercan Babakiray, Trent A. Johnson, Mikel B. Holcomb, and David Lederman

*Department of Physics, West Virginia University, Morgantown, WV 26506-6315, USA*

Transient reflectivity measurements of thin films, ranging from 6 to 40 nm in thickness, of the topological insulator $Bi_2Se_3$ revealed a strong dependence of the carrier relaxation time on the film thickness. For thicker films the relaxation dynamics are similar to those of bulk $Bi_2Se_3$, where the contribution of the bulk insulating phase dominates over that of the surface metallic phase. The carrier relaxation time shortens with decreasing film thickness, reaching values comparable to those of noble metals. This effect may result from the hybridization of Dirac cone states at the opposite surfaces for the thinnest films.

Topological insulators (TIs) are novel electronic materials that have an insulator-type band gap in the bulk (for $Bi_2Se_3$ $E_g$ ~ 0.3 eV) but have protected gapless conducting phase on their surface due to the combination of spin-orbit interactions and time-reversal symmetry.[1,2] The most effective experimental methods currently used to monitor metallic two-dimensional (2D) Dirac surface states (SS) of TIs are angle-resolved photoemission spectroscopy (ARPES) and time-resolved ARPES (TrARPES).[1-7] These techniques are equally sensitive to SS and the bulk atoms residing in the close proximity to the surface as a consequence of the extremely small penetration depth (a few nm) of incident energetic photons used for photoemission, combined with the limited escape depth of the electrons (also a few nm). Finite-size effects have also been studied for thin $Bi_2Se_3$ films of only a few nm thick and a crossover of the three-dimensional (3D) TI $Bi_2Se_3$ to the 2D limit (gapped SS) has been observed when the thickness is below six quintuple layers (~ 6 nm).[8]

Reaching a similar sensitivity to SS using traditional optical pump-probe techniques (like transient reflectivity (TR)/transmission), which use less energetic photons in the visible/infrared range, seems problematic since the absorption length of the laser light normally used for these measurements (a few tens of nm) significantly exceeds the range where the effect of SS can actually be monitored. As a result, for bulk single crystals of $Bi_2Se_3$ the transient optical response is dominated by the bulk contribution. To overcome the problem one can use SS/surface sensitive methods. An example of this approach has recently been demonstrated by illuminating $Bi_2Se_3$ with circularly polarized near-infrared light.[9] The resulting photocurrent which reverses its direction with a reversal of the helicity of the light unambiguously proves the SS origin of the optical response. Another surface sensitive technique exploits the centrosymmetric nature of TI's, which governs exclusively the surface-related response which results in an optical second harmonic generation (SHG) process.[10,11]

In this Letter we report on a new way to distinguish between the contributions from the TI ($Bi_2Se_3$) bulk 3D states and the 2D gapless SS, which is based on differences in the carrier relaxation rates for the insulating and metallic phases. We demonstrate that the carrier relaxation rate measured by the TR technique can be significantly enhanced by decreasing the thickness of $Bi_2Se_3$ films to the range preceding the energy gap opening for SS (40 - 6 nm).[8] This behavior indicates that there is a crossover between two carrier relaxation mechanisms associated with the polar phonon (Fröhlich) interaction in the bulk insulating phase and the electron-lattice interaction in the surface metallic phase. We suggest that this may result from the hybridization of Dirac cone states at the opposite surfaces of the thinnest films.

Experiments were performed on $Bi_2Se_3$ thin-film samples that were 6, 8, 10, 12, 15, 20, 25, 30, 35, and 40 nm thick. The films were grown on 0.5 mm $Al_2O_3$(0001) substrates by molecular beam epitaxy, with a 10 nm thick $MgF_2$ capping layer to protect against oxidation. The growth process was similar to the one describe in Ref. 12, except that a two-step process was implemented,[13] where an initial 3 nm $Bi_2Se_3$ layer was grown at 140 $^o$C and the rest of the sample was grown at 275 $^o$C. The polycrystalline $MgF_2$ capping layer was grown at room temperature without exposing the sample to atmosphere after the $Bi_2Se_3$ growth. The thin film thickness was determined from x-ray reflectivity (XRR) measurements and the surface roughness for the $Bi_2Se_3$ surface of all films, also determined from XRR, was one quintuple layer (~ 0.95 nm) or less. Reflection high energy electron diffraction and x-ray diffraction showed that the $Bi_2Se_3$ films were epitaxial in the plane and highly crystalline out of the plane of the film.

The TR measurements were performed with 100-fs pulses from a Ti:Sapphire laser with a center photon energy of 1.51 eV and a repetition rate of 80 MHz. Pump-probe geometry was employed, where the pump is at normal incidence and the probe is at an incident angle of ~15$^o$, focused through the same lens to a spot diameter of 100 μm. The reflection was collected by a Si photodiode and lock-in amplifier. The pump and probe beams were cross-polarized. Experiments were performed in air and at room temperature.

Figure 1 shows TR traces for a 10-nm film for varying pump-probe average power. The traces are negative, in accordance with absorption bleaching origin of the TR signals, and show a rise and multiple decay behavior, which can be fit by a multi-exponential rise-decay function as has been previously reported.[14] The rise-time constant of $\tau_R$ ~ 0.3 ps does not depend significantly on the laser power, indicating a weak dependence of the electron-electron scattering rate on the carrier density for highly energetic carriers ($\varepsilon = 1.51$ eV – $E_g = 1.21$ eV).[15] Alternatively, the decay-time constants $\tau_{D1}$, $\tau_{D2}$ and $\tau_{D3}$ show an increase with increasing laser power and illustrate the complicated relaxation dynamics of hot carriers photoexcited in $Bi_2Se_3$ films [Fig. 1, Inset (a)]. The shortest decay of $\tau_{D1} \sim 2.2$ ps is associated with electron-phonon relaxation, whereas the longer decays of $\tau_{D2} \sim 90 - 250$ ps and $\tau_{D3} \sim 10$ ns can be attributed to long-lived processes associated with the charge separation.

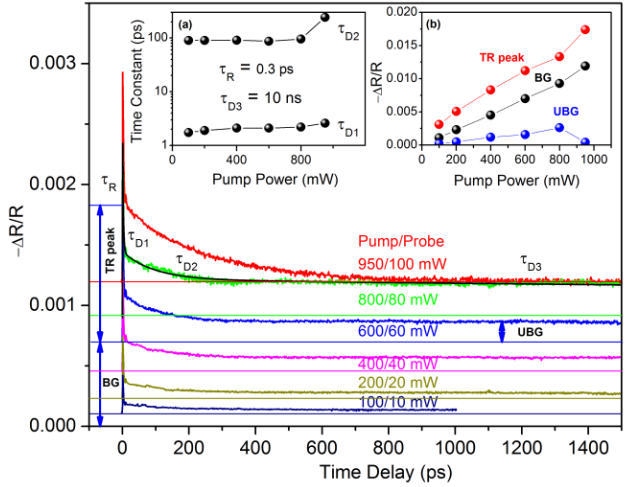

FIG. 1. TR traces for 10-nm $Bi_2Se_3$ film measured with varying pump-probe average power and the baselines for each of the curves are shown by the corresponding colors. The intensities of the BG, TR peak, and UBG signals are exampled for the 600/60 mW curve (blue). An example of the exponential fit by using rise-time constant ($\tau_R$) and three decay-time constants ($\tau_{D1}$, $\tau_{D2}$, $\tau_{D3}$) is shown as a black curve covering the 800/80 mW curve (green). Insets (a) and (b) show the pump power dependencies of the corresponding fitting parameters and signal intensities, respectively.

We also note that there is a background (BG) signal which does not depend on the delay time between the pump and the probe and is determined exclusively by the repetition rate of the laser and its average power. The BG signal is associated with a quasi-stationary population of SS at opposite surfaces of the film. Subsequently, the pump-induced population within the film continuously feeds the population of SS at the rate $1/\tau_{D2}$,[6] a process driven by a band-bending-induced space charge layer.[8] Once the process is complete, the additional pump-induced population of SS determines the unrecovered background (UBG) signal and sets up a new quasi-stationary population of SS, at a level that appears as a BG+UBG signal. Because the decay-time of the UBG signal ($\tau_{D3}$) significantly exceeds the inversed repetition rate of the laser, the films do not fully recover during the time between two sequential laser pulses (12.5 ns). This behavior implies that $\tau_{D3}$ is determined with less precision and therefore it has been set to be a constant ($\tau_{D3} = 10$ ns). The BG and UBG signals increase linearly with increasing laser power in a similar manner to the TR peak intensity [Fig. 1, Inset (b)]. At the highest laser power applied there is a deviation from linearity which mainly affects the UBG signal, reducing it to almost zero. This effect progressively increases with decreasing film thickness in the range below 12 nm (not shown explicitly) and together with a significant increase of $\tau_{D2}$ constant [Fig. 1, Inset (a)] indicates the suppression of a relaxation channel through which the pump-induced carrier population feeds the population of SS. This behavior suggests the existence of highly energetic states in the band structure of the film which can be populated mainly at high laser powers. We note that this model is qualitatively different from that employing the bulk conduction band-to-SS scattering.[6]

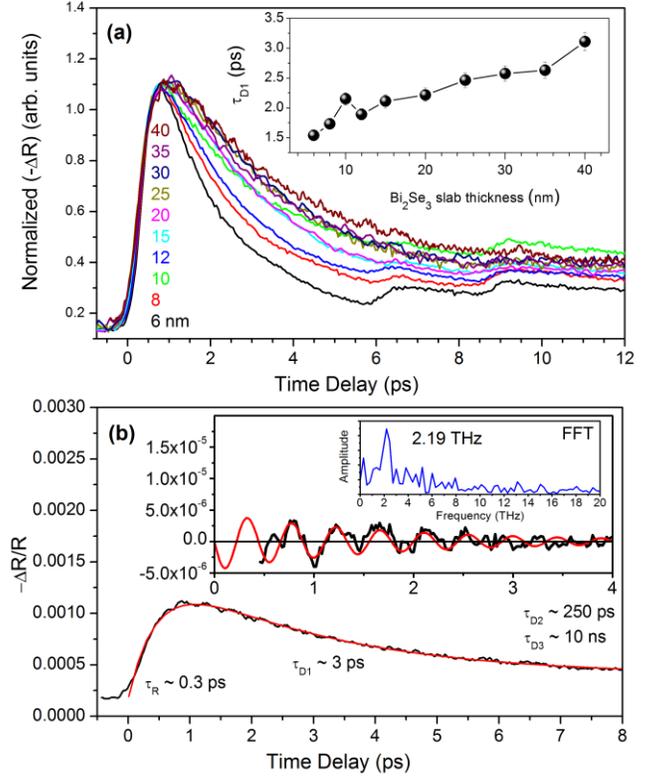

FIG. 2. (a) Normalized TR traces for $Bi_2Se_3$ films of different thickness indicated by the corresponding colors. All curves have been taken by the same Pump-to-Probe power (950/150 mW). The stepwise features between 6 and 12 ps for thinnest films result from the probe beam reflection from the 0.5 mm $Al_2O_3$(0001) substrate. Inset shows the thickness dependence of the decay-time constant $\tau_{D1}$. (b) An example of the exponential fit of the TR signal from 40-nm $Bi_2Se_3$ film. Inset shows the extracted oscillatory part of the signal (in black), a best fit by a damped cosine function with a damping coefficient of ~1.8 ps (in red), and the corresponding fast Fourier transformation (FFT) (in blue), indicating the center frequency of the mode at 2.19 THz.

Figure 2(a) shows a significant monotonic decrease of $\tau_{D1}$ constant with decreasing film thickness except for a resonance-like feature for the 10-nm sample. For thicker films (40 nm) the decay-time constant of $\tau_{D1} = 3.11$ ps agrees well with those previously reported for $Bi_2Se_3$ single crystals.[11,16,17] Furthermore, the thick films have an oscillatory behavior in the TR signals, which can be extracted and fit by a damped cosine function with a frequency of 2.19 THz [Fig. 2(b)]. The resulting frequency corresponds to longitudinal optical (LO) phonons excited in bulk $Bi_2Se_3$,[11,16,17] implying that the cooling of photoexcited hot carriers in thick films occurs through the electron-LO-phonon scattering (Fröhlich interaction). Once the thickness of films decreases, the oscillatory part of the TR signals disappears and $\tau_{D1}$ approaches the minimal value of ~ 1.5 ps (6-nm film) [Inset in Fig. 2 (a)], which is similar to the characteristic time ($\leq$ 1 ps) of internal thermalization of a nonequilibrium electron population in noble metal thin films through the electron-lattice interactions.[18] Therefore both the decrease of $\tau_{D1}$ and the disappearance of LO-phonon oscillations with decreasing film thickness result from a crossover of the carrier relaxation mechanism from one dominated by the bulk insulating phase to another one dominated by a surface metallic phase, for which the polar phonon modes do not exist.

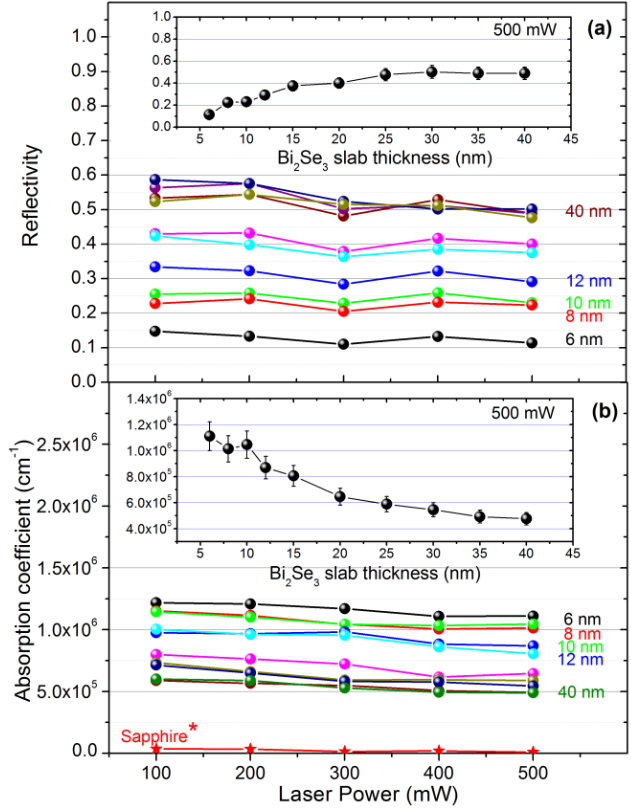

Fig. 3. Power dependencies of reflectivity (a) and absorption coefficient (b) at an incident angle of ~15° for $Bi_2Se_3$ films of different thickness indicated by the corresponding colors. Insets in (a) and (b) example the film thickness dependence of reflectivity and absorption coefficient measured with 500 mW laser power. The sapphire substrate reflectivity and absorption have been extracted. The curve marked by stars in (b) presents the absorption of the sapphire substrate.

The rate of emission of LO-phonons by hot carriers in $Bi_2Se_3$ can be estimated from the polar Fröhlich interaction given by[19]

$$\frac{1}{\tau_{e-ph}} = \frac{e^2}{4\pi\varepsilon_0\hbar}\left(\frac{2m_e^*\hbar\omega_{LO}}{\hbar^2}\right)^{1/2}\left(\frac{1}{\varepsilon_\infty} - \frac{1}{\varepsilon_s}\right),$$

where $\tau_{e-ph}$ is the characteristic electron-phonon interaction time, $e$ is the electron charge, $\varepsilon_0$ is the permittivity of free space, $\hbar\omega_{LO}$ = 8.8 meV is the LO-phonon energy, $m_e^* = 0.14 m_0$ is the electron effective mass at the $\Gamma$ point ($m_0$ is the free-electron mass), and $\varepsilon_\infty$ = 9 and $\varepsilon_s$ = 100 are the high-frequency and static dielectric constants, respectively. Consequently, the time required to emit a single LO-phonon by a hot electron can be estimated to be as long as $\tau_{e-ph}$ = 31 fs. Because the effective mass for holes in $Bi_2Se_3$ is similar to that of electrons,[20] the time required to emit a single LO-phonon by hot holes is expected to be of the same order of magnitude.

The estimation of electron-phonon coupling strength in $Bi_2Se_3$ depends on the measurement method. For example, the recently reported electron-phonon coupling constant ($\lambda$ ~ 0.08) measured using ARPES points to the exceptionally weak electron-phonon coupling regime.[21] Because this value is similar to typical values of good conductors such as noble metals ($\lambda$ ~ 0.1),[22] it can be associated with the surface metallic phase of TIs. Alternatively, the value of $\lambda$ ~ 0.43 deduced from inelastic helium-atom scattering (Ref. 23) points to the extremely strong electron-phonon coupling regime being associated with electron-polar-phonon (Fröhlich) coupling in the bulk of $Bi_2Se_3$. The relaxation of optically excited hot carriers in the strong electron-LO-phonon coupling regime is known to occur through the LO phonon cascade.[24] The excess electron energy of $\varepsilon$ = 1.21 eV used in our experiments is equivalent to ~ 137 times the LO phonon energy. The decay time $\tau_{D1}$ = 3.11 ps observed for the 40-nm film leads to an average cooling time of 23 fs per LO phonon, which agrees well with the value of 31 fs calculated above from the Fröhlich interaction, proving validity of the model.

To further understand the ultrafast carrier dynamics, we also measured the probe beam transmission and reflection for the films using the same experimental conditions as for the pump-probe measurements (Fig. 3). The reflectance ($R = I_R/I_0$, where $I_R$ and $I_0$ denote the intensities of reflected and incident beams, respectively) as a function of film thickness and laser power indicates the linearity of the excitation process. The reflectivity increases with film thickness and saturates at about $R$ = 0.5 for thicker films. This value is somewhat larger than that acquired by performing spectroscopic ellipsometry measurements of bulk $Bi_2Se_3$ (~ 0.4).[25] This deviation could result from the different incident angles and light intensities used. The absorption coefficient can be estimated from $\alpha = (1/d)\ln[I_0(1-R)/I_T]$, where $I_T$ is the intensity of transmitted light and $d$ is the film thickness in cm. The absorption coefficient dependence on laser power and film thickness is shown in Figure 3(b). The absorption coefficient for thicker films (40 nm) of $4.76\times10^5$ cm$^{-1}$ closely matches that known value for bulk $Bi_2Se_3$ ($2\times10^5$ cm$^{-1}$),[6] being increased by more than a factor of two with decreasing film thickness. Owing to the extremely large absorption coefficient of $Bi_2Se_3$ films, the corresponding light penetration depth ($1/\alpha$) becomes extremely small for the thinnest films [Fig. 4(a)], reaching values which are typical for metal layers (< 20 nm). On the other hand, the light penetration depth for the 40-nm film of ~ 21 nm agrees well with results from spectroscopic ellipsometry measurements of bulk $Bi_2Se_3$ (~ 25 nm).[25]

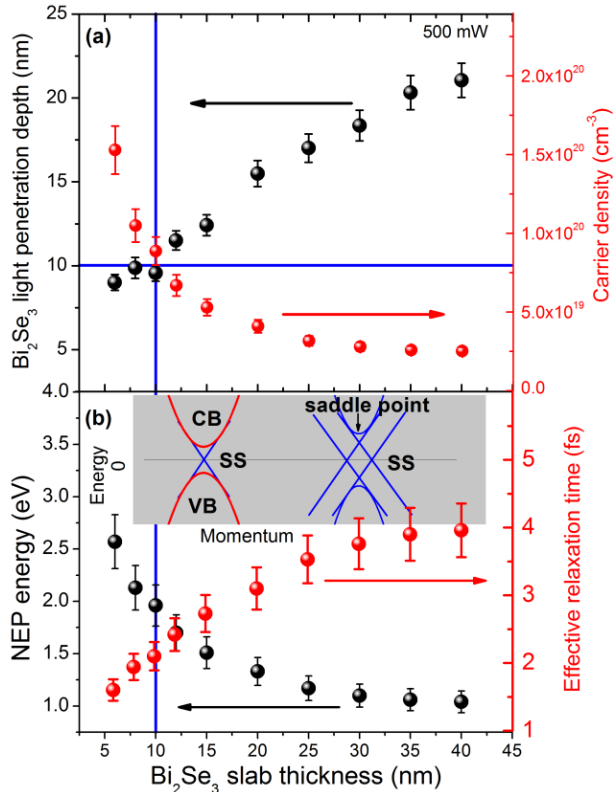

Fig. 4. (a) $Bi_2Se_3$ slab thickness dependencies of the light penetration depth (black dots) and the photoexcited carrier density (red dots) for 500 mW laser power. (b) $Bi_2Se_3$ slab thickness dependencies of non-equilibrium plasmon energy (black dots) and carrier effective relaxation time (red dots). Inset in (b) shows two model pictures representing the single Dirac cone SS of TIs and coupled Dirac cone SS.

The density of carriers photoexcited by the laser pulse in $Bi_2Se_3$ films along the $z$ direction can be estimated by using the continuity equation for the power flow. The resulting power density absorbed in the media during the laser pulse duration ($\tau_L$) is $P = -\nabla_z I$, with a laser power density $I = I_0(1-R)\exp(-\alpha z)$, where $I_0$ is the maximal laser power density at the sample surface (for example, it is 0.8 GW/cm$^2$, which corresponds to 0.08 mJ/cm$^2$ laser fluence and equally corresponds to 500 mW of averaged laser power of Ti:sapphire laser light with 80 MHz repetition rate, $\tau_L$ = 100 fs, and a focused spot diameter of ~ 100 μm); $R$ = 0.31 and $\alpha$ = $1.05 \times 10^6$ cm$^{-1}$ for 10-nm film. Subsequently, the power density absorbed within the absorption length ($d_a = 1/\alpha$) for photon energy $\hbar\omega$ = 1.51 eV is $P = \alpha I_0(1-R)\exp(-\alpha d_a)$ = $2.15 \times 10^{14}$ W/cm$^3$. The resulting average density of photoexcited carriers within the absorption length is therefore $n = (P\tau_L/\hbar\omega) \sim 8.9 \times 10^{19}$ cm$^{-3}$. Figure 4(a) shows a significant increase of photoexcited carrier density with decreasing film thickness. This tendency indicates again that the $Bi_2Se_3$ films become "more metallic" with decreasing thickness. The corresponding thickness dependence of the non-equilibrium plasmon (NEP) energy $(\hbar\omega_p)$, where $\omega_p = \sqrt{ne^2/\varepsilon_\infty\varepsilon_0 m*}$ is the plasma frequency, and the carrier effective relaxation time $(\tau_e = 2\pi/\omega_p)$ are shown in Fig. 4(b). We note here that NEP energy is shifted towards the short-wave spectral range with decreasing film thickness, similarly as that occurs in noble metal nanostructures.[26] Because the NEP energy at some thickness of the $Bi_2Se_3$ films can match closely the photon energy applied, one might expect the existence of the resonance-like features in the carrier relaxation dynamics mentioned above [Inset in Fig. 2 (a)]. A detailed discussion of these resonances is out of the scope of this paper. Also the carrier effective relaxation time confirms the reasonable assumption that the carrier relaxation rate in the metallic phase of TIs is significantly higher to that in the bulk insulating phase.

The increase of the absorption coefficient with decreasing film thickness unambiguously proves the existence of a crossover from behavior dominated by the bulk $Bi_2Se_3$ properties to phenomena dominated by the surface metallic phase as the film thickness is decreased. The crossover can be explained by a hybridization of two intrinsic Dirac states at opposite surfaces of the films. This type of coupling between Dirac cones has been recently demonstrated using ARPES for the extrinsic Dirac states of a Bi bilayer and the intrinsic surface Dirac states of $Bi_2Te_3$ film which are close in energy.[27] The inset in Figure 4(b) illustrates the single Dirac cone TI states and coupled Dirac cone TI states originating from the opposite surfaces of the thinnest $Bi_2Se_3$ films. The existence of Dirac cone hybridization has also been theoretically predicted for bilayer graphene, where the saddle point between split Dirac cones in the band structure is attributed to a van Hove singularity in the density of states.[28] The highly energetic states discussed above might be associated with the saddle point states between Dirac cones at opposite surfaces of the films. Despite the fact that the existence of these coupled states and optical transitions between them are currently controversial topics, their fully metallic nature seems to be beyond question.

In summary, we have provided experimental evidence that ultrafast carrier dynamics in TI $Bi_2Se_3$ films between 6 and 12 quintuple layers thick are predominantly governed by the properties of the surface metallic phase, which dominates over the bulk insulating phase. We suggest that the existence of Dirac cone coupled states originating from the opposite surfaces of the ultra-thin $Bi_2Se_3$ films can possibly explain the experimental observations.

We thank A. Bristow and P. Borisov for useful discussions. This work was supported by a Research Challenge Grant from the West Virginia Higher Education Policy Commission (HEPC.dsr.12.29). Some of the work was performed using the West Virginia University Shared Research Facilities.